\documentclass[
aps,superscriptaddress,
twocolumn,prd,
showpacs,preprintnumbers,nofootinbib,
amsmath,amssymb,
aps,
superscriptaddress]{revtex4-1}
\usepackage{Preamble}
\usepackage{colortbl}
\graphicspath{{./figures}}
\makeatletter
\AtBeginDocument{\let\includegraphics\saved@includegraphics}

\newcommand{\cor}[1]{{\textcolor{black}{#1}}}
\newcommand{\ncor}[1]{{\textcolor{black}{#1}}}
\newcommand{\nncor}[1]{{\textcolor{darkred}{#1}}}

\begin{document}

\title{A complete measurement of a black-hole recoil through higher-order gravitational-wave modes}

\author{Juan Calder\'on~Bustillo}
	\affiliation{Instituto Galego de F\'{i}sica de Altas Enerx\'{i}as, Universidade de
Santiago de Compostela, 15782 Santiago de Compostela, Galicia, Spain}
	\affiliation{Department of Physics, The Chinese University of Hong Kong, Shatin, N.T., Hong Kong}

\author{Samson H. W. Leong}
	\affiliation{Department of Physics, The Chinese University of Hong Kong, Shatin, N.T., Hong Kong}
	
\author{Koustav Chandra}
	\affiliation{Department of Physics, Indian Institute of Technology Bombay, Powai, Mumbai, Maharashtra 400076, India}
     \affiliation{Institute for Gravitation and the Cosmos, Department of Physics, Pennsylvania State University, University Park, PA 16802, USA}

\begin{abstract}
General relativity predicts that gravitational waves (GWs) carry linear momentum. Consequently, the remnant black hole of a black-hole merger can inherit a recoil velocity or ``kick'' of crucial implications in, \textit{e.g.}, black-hole formation scenarios. While the kick magnitude is determined by the mass ratio and spins of the source, estimating its direction requires a measurement of the \textit{two orientation angles} of the source. While the orbital inclination angle is commonly reported in GW observations, the scientific potential of the azimuthal one has not been exploited to date. We show how the presence of more than one GW emission mode allows one to constrain this angle and, consequently, the kick direction of a real GW event. We analyse the GW190412 signal, which contains higher-order modes, with a numerical-relativity surrogate waveform model for black-hole mergers. We rule out kick magnitudes below the typical escape velocity of dense globular clusters $v_{\text{esc}}\approx 50$\,km/s with a Bayes Factor of $\simeq 21$ (or $\simeq 95\%$ probability). The kick forms angles $\theta_{KL}^{-100M}=32^{+35}_{-14}\,\deg$ with the orbital angular momentum defined at a reference time $t_{\rm ref}=-100\,M$ before merger (with $M$ denoting the system mass in geometric units), $\theta_{KN}=44^{+19}_{-17}\,\deg$ with the line-of-sight. The projections of the kick and line-of-sight onto the orbital plane form an angle $\phi_{KN}^{-100M}=69^{+33}_{-38}\,\deg$. All quantities are quoted at a $90\%$ credible level. Finally, by analyzing numerically simulated signals, we show that recoils can be estimated in an unbiased way using the NRSur7dq4 waveform model. We briefly discuss the potential application of this type of measurement for multi-messenger observations of black-hole mergers occurring in Active Galactic Nuclei.

\end{abstract}

\maketitle

\section{Introduction} Gravitational waves (GWs) carry linear momentum away from their sources \citep{Misner:1974qy} and asymmetric black-hole (BBH) mergers emit GWs in an anisotropic way. This causes a net emission of linear momentum that makes the final black hole (BH) acquire a recoil velocity~\citep{Thorne:1980ru, 1983MNRAS.203.1049F, maggiore2008gravitational, Gonzalez:2006md, Herrmann:2007ac, Koppitz:2007ev, Sundararajan:2010sr, Lousto:2011kp, CalderonBustillo:2018zuq}, or kick, that in the most extreme cases can reach ${\cal{O}}(1000)$\,km/s~\citep{Campanelli:2007cga, Gonzalez:2007hi, Brugmann:2007zj, Healy:2008js, Sperhake2020_ecc, Lousto:2011kp,Lousto:2012gt,Lousto_wet_merger_kick}. Such speeds can expel the remnant BH from their host environments preventing them from taking part in subsequent mergers and, therefore, from contributing to hierarchical BH formation mechanisms~\citep{GerosaFishbach21}. This has crucial astrophysical consequences as, {\it e.g.}, such scenarios \cor{can explain} the formation of supermassive BHs \cite{Volonteri2003,Volonteri2010,Sakurai2017}. Due to its paramount importance in BH formation, much work has been devoted towards estimating the kick magnitude of the remnant BHs of the GW events observed by Advanced LIGO~\citep{LIGOScientific:2014pky} and Advanced Virgo~\citep{VIRGO:2014yos}~\citep{CalderonBustillo:2018zuq, Vijay_kick, GW190521I, Vijay_GWKick,Kick_mahapatra, Islam_SurrCatalog}. The kick magnitude, together with its direction in the source frame, is determined by the mass ratio and spins of the merging BHs \cite{Campanelli:2007cga, Gonzalez:2007hi, Brugmann:2007zj, Healy:2008js, Lousto:2011kp,Lousto:2012gt}, with the spin relative orientations playing a particularly important role. Estimating these, however, \cor{is intrinsically challenging}~\citep{Purrer:2015nkh, Biscoveanu:2021nvg}. Consequently, only \cor{three} of the $\sim \mathcal{O}(100)$ events detected to date~\citep{LIGOScientific:2018mvr, Nitz:2018imz, Venumadhav:2019lyq, Nitz:2020oeq, LIGOScientific:2020ibl, GWTC3, Nitz:2021uxj, Nitz:2021zwj, Olsen:2022pin}, namely GW190814~\citep{Measured_kick_magnitude_GW190814}, \ncor{GW191109\_010717 \cite{Islam_SurrCatalog}} and GW200129\_065458~\citep{Vijay_GWKick} (GW200129 hereafter), have allowed for an informative estimation of the recoil magnitude.
Estimating the recoil direction with respect to the observer requires, in addition, an estimate of the two angles characterising the Earth's location on the binary's sky, which can be conversely understood as the BBH orientation w.r.t.~the observer. We can access such information through the gravitational-wave mode content of the signal, as we outline next.

\begin{figure}[h!]
    \centering
    \includegraphics[width=0.9\columnwidth]{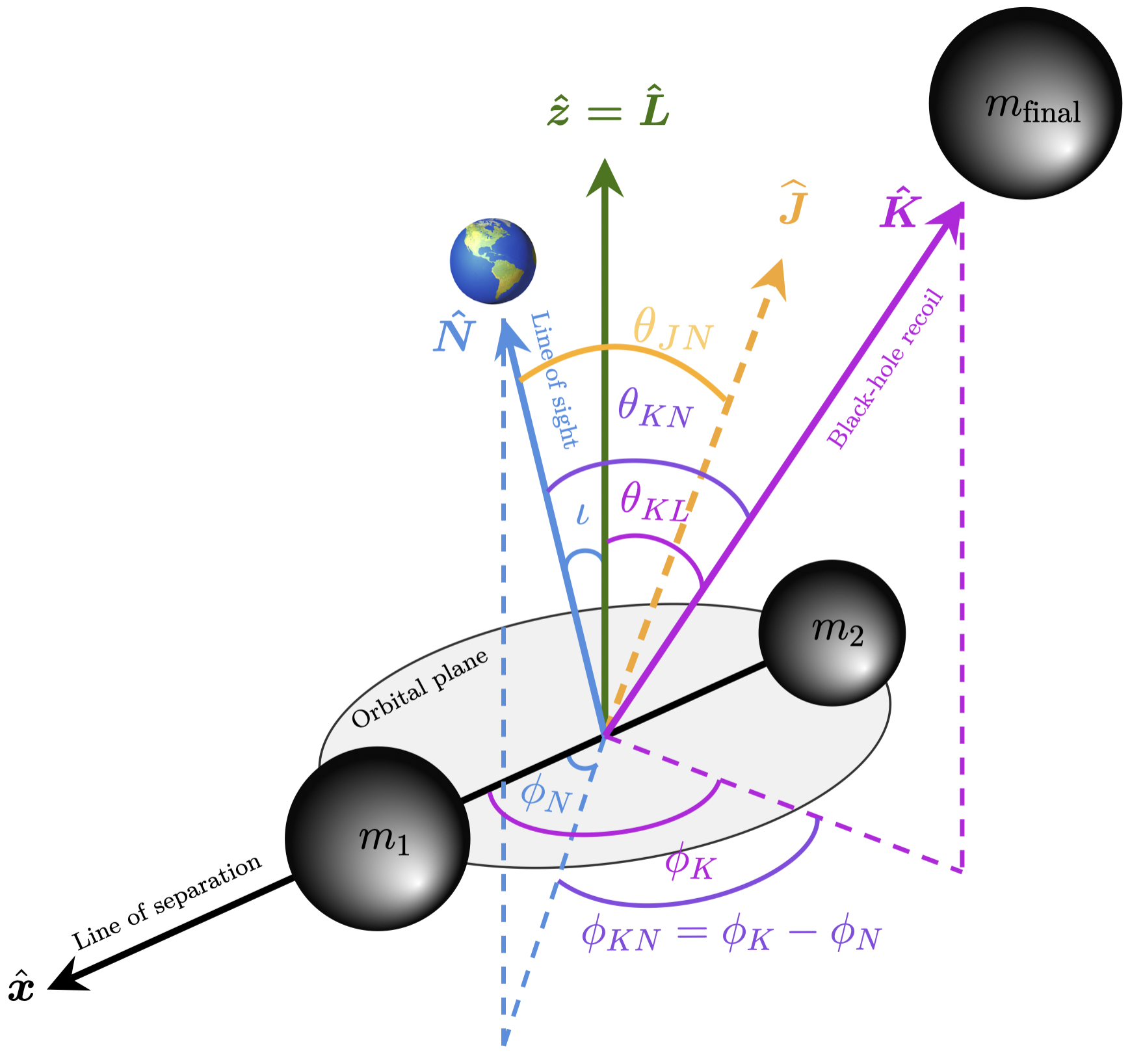}
    \caption{\textbf{Sketch of our black-hole merger reference frame} The polar and azimuthal angles $(\iota,\,\phi_N)$ characterise the orientation of the orbital plane of a black-hole merger or, conversely, the direction of the line-of-sight $\vu*{N}$ on its sky. The vector $\vu*K$ represents the final black-hole recoil (or kick). Its direction on the binary's sky is characterised by $\theta_{KL}$ and $\phi_{K}$. Finally, we characterise its direction with respect to the line-of-sight by the angles $\theta_{KN}$ and $\phi_{KN}$.}
    \label{fig:kick}
\end{figure}

\section{Methods}

\subsection{The source frame and the orientation angles} The GW emission from a BBH can be described as a superposition of GW modes $h_{\ell,m}$ \cor{multiplied by spin\nobreakdash\ncor{(-2)}-weighted spherical harmonics} $Y^{-2}_{\ell,m}$ as~\citep{Goldberg:1966uu, Blanchet2024}:
\begin{equation}
    h_+ - i h_\times = \sum_{\ell,m} Y^{-2}_{\ell,m}(\iota,\phi_N)\, h_{\ell,m} (\Xi;t)\,.
\end{equation}
Here, $\Xi$ denotes the intrinsic parameters (masses and spins) of the source. The parameter $\iota$ denotes the polar angle formed between the line-of-sight (LOS) and the (instantaneous) \cor{Newtonian} orbital angular momentum $\vb*{L}$, normal to the orbital plane defined at a given reference time $t_{\rm ref}$ during the BBH evolution. The value $\iota=0$ denotes a ``face-on'' observer while $\iota=\pi/2$ denotes an ``edge-on'' observer located on the orbital plane. The angle $\phi_N$ denotes the azimuthal angle of the observer, {\it i.e.}, the angle formed by the projection of the line-of-sight $\vb*{N}$ onto the orbital plane and some preferential axis $\vu*{x}$~\citep{CalderonBustillo:2018zuq} on it, which is commonly chosen\footnote{We note that waveform models used within the LIGO, Virgo and KAGRA (LVK) Collaborations, implemented within the software \texttt{LALSuite}~\citep{lalsuite}, use a definition for the azimuthal angle $\phi_N^{\text{LVK}}=\pi/2-\phi_N$. (See \texttt{phiRef} or $\Phi$ in \citep{Schmidt:2018btt}.)} as the vector pointing from the lighter to the heavier BH~\citep{Schmidt:2018btt}.\\ 

\subsection{Characterising the kick direction and re-defining the azimuthal angle} Within the above frame, the final BH kick $\vb*{K}$ can be characterised by its magnitude $K$ and the angles $(\theta_{KL},\,\phi_{K})$ that it forms with $\vb*{L}$ and $\vu*{x}$ (See Fig.~\ref{fig:kick}). Using the kick and the LOS, we can compute two more ``observer-related'' angles: the angle $\theta_{KN}$ subtended between them and the angle $\phi_{KN}=\phi_K-\phi_N$ formed by their projections on the orbital plane.\\ 

BBHs with spins (anti-)aligned with $\vu*{L}$ display a constantly oriented orbital plane. In this case, although $\iota$ is time-independent, $\phi_N$ clearly depends on $t_{\rm ref}$. This partly motivates the latter to be systematically treated as a sort of ``nuisance'' parameter not reported in GW catalogues, commonly referred to as ``coalescence phase''. The angle $\phi_{KN}$, however, provides a time-independent and more astrophysically motivated re-definition of the azimuthal location of the observer~\cite{CalderonBustillo:2018zuq} which, in the following, we will use to characterise both the source orientation and the kick direction.\\

Finally, for generically spinning BBHs, spin-orbit coupling causes $\vb*{L}$ to precess around the total angular momentum $\vb*{J}$ \cite{Apostolatos:1994mx}\footnote{Throughout this work, whenever is needed, we compute $\vb*L$ from the component spin vectors and orbital frequency up to 3.5~post-Newtonian order~\cite{Bohe2013:3p5PN,Blanchet2024}, including \nncor{spin-orbit effects}.}. This leads to a time-dependent orientation of the orbital plane, $\iota$ and $\phi_{KN}$. In this situation, it is common to replace the inclination angle $\iota$ by the angle $\theta_{JN}$ \cite{Farr2014} formed between the LOS and the almost-conserved direction of $\vb*{J}$ (see Fig.~\ref{fig:kick}). Analogously, we can replace $\phi_N$ by the angle $\phi_N^J$ formed between the projections of the LOS and $\vb*{K}$ onto the plane normal to $\vb*{J}$.\\

\subsection{Reading the kick and orientation angles from gravitational waves: GW200129 and GW190412} The spherical harmonics $Y^{-2}_{\ell,m}(\iota,\phi_N)$ can be decomposed into amplitude and phase terms as $Y^{-2}_{\ell,m}(\iota,\phi_N) = |Y^{-2}_{\ell,m}(\iota,0)|\,\eu^{-\iu m\phi_N}$. This shows that while $\iota$ controls the amplitude of each mode, $\phi_N$ determines the relative phase with which these modes interact, dramatically impacting the morphology, ({\it i.e.} the frequency content) of the observed signal~\citep{CalderonBustillo:2019wwe}. \footnote{\textcolor{black}{Strictly, one needs either orbital precession or the observation of a higher-mode with odd-$m$ --suppressed for equal-mass BBHs-- which leads to a non-periodic signal as a function of $\phi_N$. For instance, the observation of e.g. a $(4,4)$ mode in an equal-mass non-precessing binary would lead to a bi-modal distribution for $\phi_N$ due to the periodicity of the signal as a function of $\phi_N$.}}. Consequently, $\phi_N$ \textcolor{black}{should be measurable if two modes with distinct frequency content are observed in the signal. This, however, requires the observation of either a precessing and/or unequal-mass BBH with non-zero inclination, which is challenging} 

\textcolor{black}{First, the BBH emission is vastly dominated by the so-called ``quadrupole'' modes $(\ell,\,|m|)=(2, 2)$ while further modes, known as higher-order modes (HMs), only contribute significantly for asymmetric sources with orientations $\iota \neq (0,\pi)$. Second, while the joint observation of both dominant modes permits the estimation of $\phi_N$ for precessing sources~\cite{OShaughnessy2013}, these are however related by $h_{2,-2}=h^{*}_{2,2}$ for non-precessing ones, reducing the impact of $\phi_N$ to a trivial phase shift ~\cite{Blanchet2024, Harry:2016ijz, Harry:2017weg} that prevents its measurement. Therefore, for non-precessing cases we can only measure $\phi_N$ through the observation of HMs with $m\neq 2$, which requires unequal-mass sources that are not face-on/off.}

\begin{figure*}
\begin{center}
\includegraphics[width=0.99\textwidth]{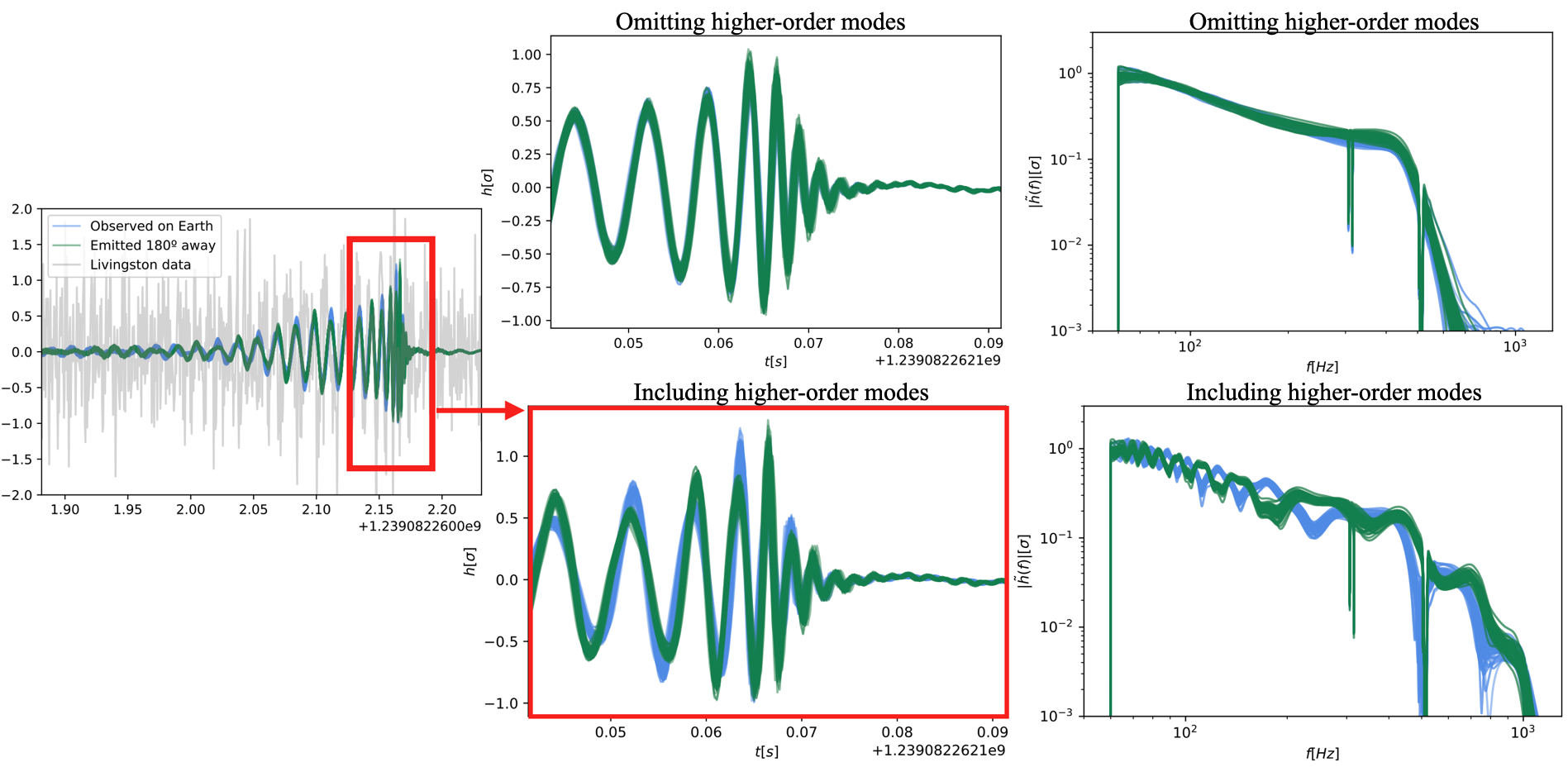}
\caption{\textbf{GW190412 as observed on Earth and 180 deg away: impact of higher-order modes.} The left panel shows the whitened LIGO Livingston data around GW190412 together with the corresponding top-100 highest likelihood (best fit) waveforms in blue. In green, we also show the signals emitted by the source in the direction opposite to the line-of-sight, which clearly differ. The mid-bottom panel zooms into the late-inspiral and merger region to highlight morphological differences. Finally, the right-bottom panel shows the corresponding Fourier transforms. The top central and right panels show the same as the bottom ones, but restricting the waveforms to only the dominant quadrupole $(2,\pm 2)$ modes. This removes the information about the azimuthal angle, making the two sets of waveforms indistinguishable.}
\label{fig:waveforms}
\end{center}
\end{figure*}

In addition, detecting these types of signals is nowadays very challenging. First, signals from highly inclined systems are \cor{weaker} than those emitted face-on, reducing the chance of detection. Second, current matched-filter searches only target the quadrupole modes of non-precessing BBHs, \cor{reducing} our sensitivity towards HM-rich or precessing signals~\citep{Capano:2013raa, Varma:2014jxa, Bustillo:2016gid, Harry:2016ijz, Harry:2017weg, CalderonBustillo:2017skv, Chandra:2020ccy, Chandra:2022ixv,Schmidt_prec_gstlal,Schmidt_prec_gstlal_assym}.

Despite this, the LVK confidently observed three such signals during its third observing run. First, GW200129~\citep{GWTC3} displays signatures of precession~\citep{Hannam_nature_precession}, allowing for an informative estimation of the kick magnitude $K>698$\,km/s~\citep{Vijay_GWKick} and, although not emphasised, a certain characterisation of the kick direction \cor{(see their Fig. 3)}.
The interpretation of GW200129, however, has been challenged by~\citet{Curious_GW200129}, due to potential data-quality issues in the Livingston detector\footnote{See however \cite{Macas2024}.}. Second, while GW190814~\cite{GW190814} contains HMs and enabled a kick-magnitude measurement \cite{Measured_kick_magnitude_GW190814}, its mass ratio exceeds the limits of existing waveform models directly calibrated to numerical simulations \cite{NRSur7dq4,Varma2019_hyb8}, presumably needed to accurately estimate the kick direction. With this, we focus on the BBH event GW190412~\citep{GW190412} which, while showing no signatures of precession, has a mass-ratio $q\simeq 3$ and orbital inclination $\iota\geq 30\,\deg$ at the $90\%$ credible level. Consequently, GW190412 contains measurable HMs \cite{Roy2021_HMs}, making it a \cor{suitable} candidate for constraining the kick direction.

\subsection{Parameter inference} We perform Bayesian parameter inference on 4 seconds of publicly available data from the two Advanced LIGO and the Advanced Virgo detectors around the time of GW190412, sampled at 2048\,Hz, using the software \texttt{Parallel Bilby}~\cite{Ashton:2018jfp,Smith:2019ucc}. We compare GW190412 with the state-of-the-art BBH waveform template model \texttt{NRSur7dq4}~\citep{NRSur7dq4}, which includes the impact of orbital precession and HMs.\\ 

\begin{figure*}
\begin{center}
 \includegraphics[width=0.36\textwidth]{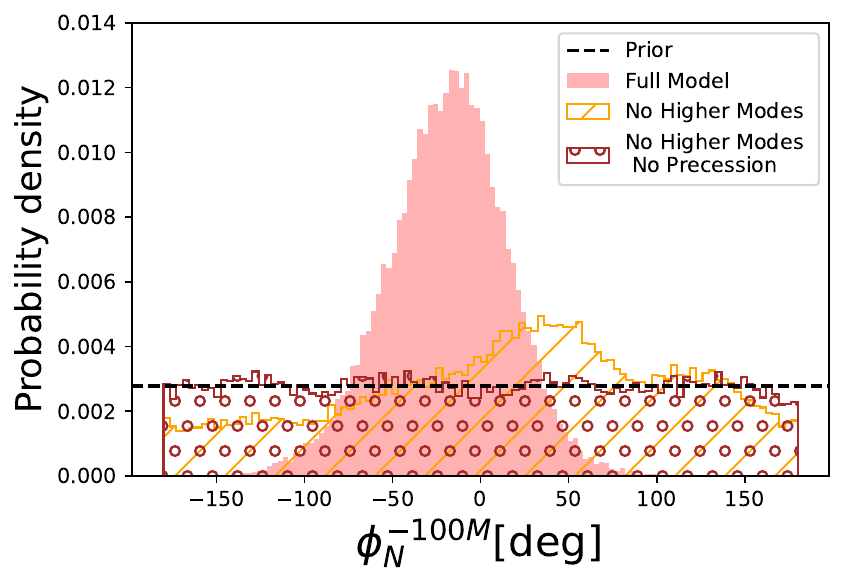}
 \includegraphics[width=0.312\textwidth]{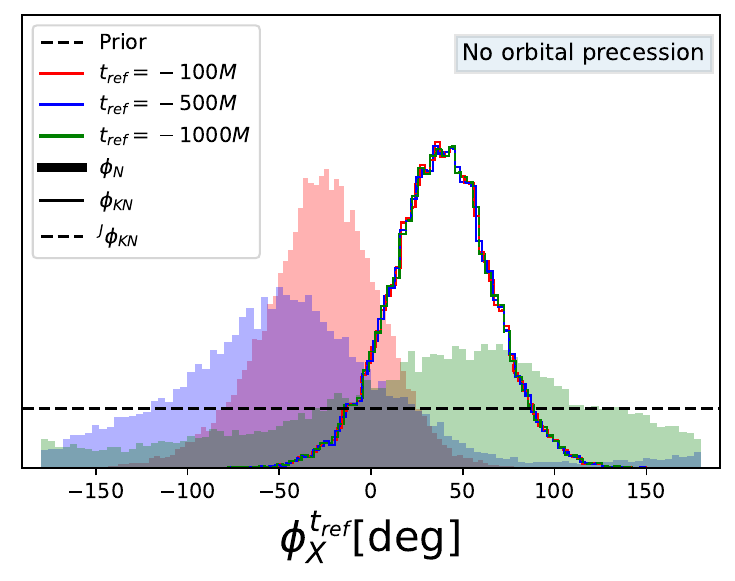}
 \includegraphics[width=0.312\textwidth]{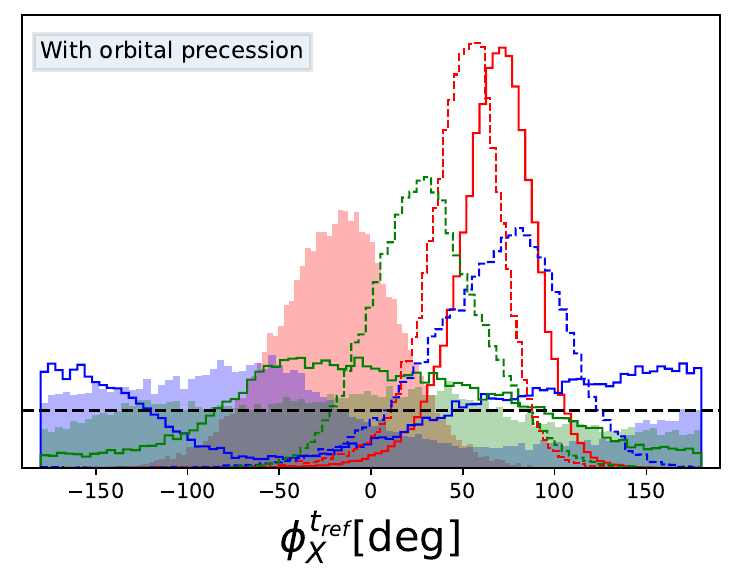}
\caption{\textbf{Azimuthal angle around GW190412: impact of higher-order modes and precession}. The left panel shows the posterior distributions of the azimuthal location of Earth $\phi_N^{-100\,M}$ around GW190412, defined as the angle between the projection of the line-of-sight onto the orbital plane and the line joining the two BHs at a time $t_{\rm ref}=-100\,M$ before merger. We show this for analyses including higher modes and precession, ignoring higher modes and ignoring both effects. The filled histograms in the central and rightmost panels show the same quantity, but also computed at times $t_{\rm ref}=-500\,M$ and $-1000\,M$. The empty histograms, instead, show the angle formed by the projections of the kick and the line-of-sight on the orbital plane $\phi_{KN}^{t_{\rm ref}}$ (solid) and the plane normal to the total angular momentum $\vb*{J}$ (dashed)  $^{J}\phi_{KN}^{t_{\rm ref}}$. In the central panel, we ignore orbital precession in the analysis while in the right panel we include it. }
\label{fig:phase}
\end{center}
\end{figure*}

Our waveform model choice is motivated by two factors. 
First, unlike alternative models for precessing BBHs used by the LVK for the analysis of this signal (namely \texttt{IMRPhenomPv3HM}~\citep{PhenomPv3HM} and \texttt{SEOBNRv4PHM}~\citep{Ossokine:2020kjp,Pan:2013rra,Babak:2016tgq}), \texttt{NRSur7dq4} is directly fitted to generically precessing numerical relativity (NR) simulations. In particular, on the one hand, the first two models reproduce the impact of precession through post-Newtonian \cite{Schmidt:2012rh,Hannam:2013oca} and \cor{effective-one-body approximations} \cite{Buonanno2003_Prec} that break down near the merger stage \cite{Hannam_nature_precession,Ossokine:2020kjp}. On the other hand, and more crucial for this study, the relative phase of the individual $h_{\ell,m}$ modes of these models is not calibrated to NR during the merger-ringdown stage, which can lead to biased estimations of both the kick magnitude, direction and azimuthal angle~\cite{Angela_kick}. Second, the associated model \texttt{NRSur7dq4Remnant}~\citep{NRSur7dq4} provides accurate estimations of the magnitude and direction of the kick given the BBH parameters. As a shortcoming, the limited time length of the waveforms generated by \texttt{NRSur7dq4} prevents the analysis of the full signal. \cor{Consistently with the re-analysis of GW190412 done by \cite{Islam2021_GW190412} using \texttt{NRSur7dq4}, we start our analysis at a frequency $f_{\rm min}=40$\,Hz, instead of the value $f_{\rm min}=20$\,Hz used by the LVK. Consistently with \cite{Islam2021_GW190412}, despite the missing information in the $20-40$\,Hz band \footnote{Also, some higher-order modes in our waveforms, such as the $(3,3)$ or $(4,4)$ will start above 40\,Hz. \textcolor{black}{Using the \texttt{IMRPhenomXPHM} waveform model \cite{XPHM_Pratten}, we have checked that the corresponding missing information -- in terms of signal-to-noise ratio -- does not impact our results, consistently with \cite{Islam2021_GW190412}}.}, we obtain parameter estimates consistent with those of the LVK, plus informative estimates of both the magnitude and direction of the kick.}\\

We place uniform priors in the detector-frame masses, spin magnitudes, time-of arrival and signal polarisation, together with isotropic priors in spin and source orientation and a luminosity distance prior $\pi(d_L)\propto d_L^2$ as in \cite{GW190412,GW150914_properties}. We sample the likelihood on the parameter space using the nested sampler \texttt{Dynesty}~\citep{speagle2020dynesty} with 4096 live points \footnote{\textcolor{black}{We do not marginalise over calibration uncertainty, as this is not expected to have a significant impact at the current detector sensitivity} \cite{Vitale2012_calibration,Payne2020_calibration,Huang_calibration}.} 
Finally, in order to show that the kick direction characterisation comes from the information encoded in the HMs, we also analyse GW190412 by removing HMs and/or precession from our templates. \cor{We report our results as median values together with symmetric 90\% credible intervals.}\\

\section{Results}

\subsection{Visualising GW190412 and the kick impact} The left panel of Fig.~\ref{fig:waveforms} shows the whitened data from the Livingston detector at the time of GW190412 (grey) together with the 100 best-fitting templates (blue). In green, we show the corresponding signals observed in the opposite direction around the source, {\it i.e.}, observed at $(\iota^{\rm Earth},\,\phi_N^{\rm Earth}+\pi)$. The bottom-central panel shows the last few cycles of these waveforms while the top-central panel shows the same waveforms, but with the HMs removed. While in the first case the two sets of signals clearly differ, therefore enabling measurement of $\phi_N$ and $\phi_{KN}$, these are almost identical when HMs are removed, preventing such measurements. Finally, the rightmost panels show these waveforms in the frequency domain. The bottom panel makes it obvious that the two waveforms show very different frequency content as a result of different interactions of the GW modes \cite{CalderonBustillo:2018zuq,CalderonBustillo:2019wwe}. Finally, the central and right bottom panels show that the \cor{kick direction is encoded in morphological waveform differences arising from the varying mode combinations in the recorded by different observers around the source \cite{CalderonBustillo:2018zuq,CalderonBustillo:2019wwe}, which vastly dominate  putative Doppler shifts arising from the speed of the source relative to the observer \cite{Gerosa:2016vip}}. \\

\subsection{Earth's azimuthal angle $\phi_N$ around GW190412} 
The left panel of Fig.~\ref{fig:phase} shows in red the posterior distributions for $\phi_N$ inferred when both HMs and orbital precession are included in the analysis, estimated at a reference time $t_{\rm ref}=-100\,M$ before merger\footnote{\ncor{In geometric units, setting $G=c=1$, with $M$ denoting the total mass of the binary in geometric units.}}. The prior probability is shown in black. The two remaining posteriors omit  HMs, the circle-filled one omitting also precession. Ignoring HMs yields uninformative posteriors, as expected, while the inclusion of HMs leads to a clearly informative posterior, yielding $\phi^{-100\,M}_N = -7^{+44}_{-42}\,\deg$\footnote{We note that when precession is considered, certain information on $\phi_N$ can be retrieved even if ignoring higher modes \cite{OShaughnessy2013}. The reason is that while the $(2,\pm 2)$ modes are related by $h_{2,-2}=h^{*}_{2,2}$ for non-precessing sources, and therefore have identical frequency content, such relation does not hold in general.}. To highlight the time-dependence of $\phi_N$, the filled histograms of the central panel show the posterior for $\phi^{-100\,M}_N$ compared to those obtained at $t_{\rm ref}=-500\,M$ and $-1000\,M$ when orbital precession is omitted. 
The empty histograms, instead, show the angles $\phi_{KN}$, computed on the orbital plane (solid lines), and $^{J}\phi_{KN}$, computed on the plane normal to $\vb*{J}$ (dashed). These two are coincident and time-independent in the absence of precession, providing a clear physical interpretation of the azimuthal angle. Finally, the right panel shows the same quantities as the central one, with precession included. Estimates for $^{J}\phi_{KN}$ are more informative and stable for varying $t_{\rm ref}$ than those for $\phi_{KN}$, owing to the lower variation of $\vb*{J}$ as the system evolves. Both estimates become not only way more informative at $t=-100\,M$ -- when the gravitational field is stronger \cite{Varma_100M,Vijay_GWKick} -- but also highly consistent\footnote{\ncor{After $-100\,M$, spin measurements start to become unreliable in NR simulations \textcolor{black}{\cite{NRSur7dq4}}.}}. This indicates that $\vb*{J}$ and $\vb*L$ are rather aligned near the merger, consistently with the lack of evidence for precession in GW190412\footnote{See Ref.~\cite{GW190412} and Sec. III C.}. In the following, we quote measurements using $t_{\rm ref}=-100\,M$\footnote{We note that while we directly sample the parameter space using this $t_{\rm ref}$, we have checked that \textcolor{black}{almost identical posterior distributions} are obtained by sampling at a reference frequency of 60\,Hz, and then evolving the corresponding spin and orientation angles as predicted by both $\texttt{NRSur7dq4}$ and post-Newtonian theory, \textcolor{black}{which we use as a robustness test. For the spin evolution, we used the \texttt{SpinTaylorT5}~\cite{Ajith2011:SpinTaylorT5,SpinTaylorDCC} scheme, and adapted the implementation in \texttt{PESummary}~\cite{*[{}][{ in particular, this piece of wrapper \href{https://git.ligo.org/lscsoft/pesummary/-/blob/543d4c70a3d8d35d58e21a85411815c729af2666/pesummary/gw/conversions/evolve.py}{code}.}] Hoy2021:PESummary} to make use of reference times instead of reference frequencies. \textcolor{black}{The \texttt{PESummary} code makes use of the evolution functions from \texttt{LALSuite}~\citep{lalsuite}.}}}. \\

\begin{figure}
\begin{center}
\includegraphics[width=0.49\textwidth]{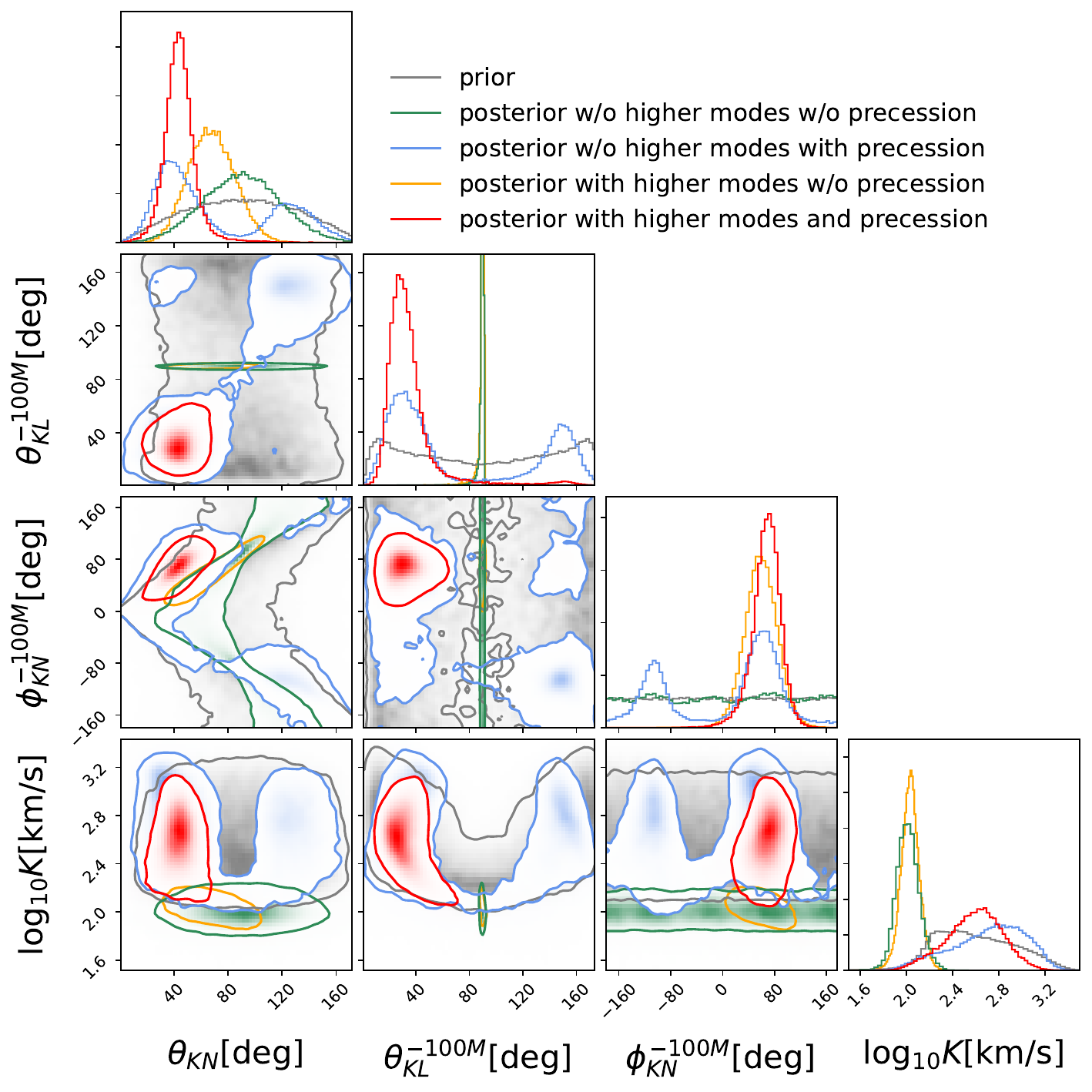}
\caption{\textbf{Magnitude and direction of the GW190412 recoil.} The side panels show the one-dimensional posterior distribution for the magnitude of the kick of GW190412 \cor{(in terms of its base-10 logarithm)}, together with those for the angles it forms with the line-of-sight $\theta_{KN}$, the orbital plane $\theta_{KL}^{-100M}$, and the projection of the former onto the latter $\phi^{-100M}_{KN}$. The orbital plane is defined at a reference time $t_{\rm ref}=-100\,M$ before merger. The inner panels show the corresponding 2-dimensional $90\%$ credible regions. We show in grey the corresponding priors \cor{for the case where precession is included.}}
\label{fig:kick_pe}
\end{center}
\end{figure}

\subsection{The kick of GW190412}

Fig.~\ref{fig:kick_pe} shows, in red, the two-dimensional $90\%$ credible regions for the kick magnitude $K$ and the angles $\theta_{KN}$, $\theta_{KL}^{-100\,M}$ and $\phi_{KN}^{-100\,M}$, together with the corresponding one-dimensional distributions. The prior distributions are shown in grey \cor{for the precessing case}. While the kick magnitude is largely unconstrained, \cor{yielding $K=400^{+509}_{-256}$\,km/s, the posterior significantly deviates from the prior, specially in the low-kick region. This allows us to rule out kicks below the typical escape velocity $v_{\rm esc}\approx 50$\,km/s \textcolor{black}{of globular clusters and young-star clusters} \cite{HolleyBockelmann2008_GlobClusterScape,Merritt:2004xa,Antonini2016,Stoop2023_esc_ygc, Stoop2023_21kms, PortegiesZwart2010, Mapelli2021_escape_vels}, which are the best-motivated environments for GW190412~\cite{Gerosa2020_GW190412}}\footnote{This is based on the assumption that GW190412 formed hierarchically. However, Ref.~\cite{Gerosa2020_GW190412} also mentions that such environments are actually unlikely, precisely due to their low escape velocities.} \cor{with a Bayes Factor of ${\cal B}\simeq 21$} or, equivalently, $\sim 95\%$ probability (see Appendix~\ref{app:BayesZ}).\\

GW190412 contains rich information about the kick direction. First, the kick forms an angle \cor{$\theta_{KN}=44^{+19}_{-17}\,\deg$} with the LOS. Removing any combination of HMs and precession (green, red, magenta) leads to less informative or biased posteriors. Second, the prior for $\theta_{KL}$ is symmetric around $90\,\deg$ ({\it i.e.}, with respect to the orbital plane) with peaks at $\simeq 20$ and $160\,\deg$. This reflects the well-known fact that precessing sources preferentially lead to kicks out of the orbital plane \cite{Brugmann:2007zj,Lousto:2012gt,SurfinBH,OShaughnessy2013}, but with no preference for shooting the final BH up or down, {\it i.e.}, with a positive or negative projection of the kick onto $\vb*{L}$. In contrast, the posterior distribution \cor{conclusively rules out a ``negative projection'', indicating that the kick formed an angle $\theta_{KL}^{-100M} = 32^{+35}_{-14}\,\deg$ with $\vb*{L}$ ({\it i.e.} $\simeq 58\,\deg$ with the orbital plane)}. Again, removing HMs leads to a less informative posterior that, in particular, barely distinguishes between the ``upper'' and ``lower'' branches. As expected, removing precession constrains the kick to the orbital plane. Third, we obtain an informative posterior for $\phi^{-100\,M}_{KN}$ yielding \cor{$\phi_{KN}^{-100\,M}= 69^{+33}_{-38}\,\deg$}. Similar to Fig.~\ref{fig:phase}, removing both HMs and precession (green) yields a completely uninformative posterior while, if precession is included, a little information is retrieved thanks to the asymmetry between the $(2,\,\pm2)$ modes (blue). Finally, we note that ignoring precession but including HMs (orange) yields results consistent with those including both effects. We understand this is consistent with the fact that orbital precession is not needed to explain GW190412~\cite{GW190412}\footnote{\cor{In fact, we obtain a Bayes factor of only 2.1 in favour of the precessing hypothesis over the non-precessing one.}}.\\

Finally, while \cite{Vijay_GWKick,SurfinBH} suggest that some kick-direction estimates of \texttt{NRSur7dq4Remnant} may be inaccurate for kicks below $\sim 300$\,km/s, we have thoroughly checked that our results are not impacted by this (see Appendix~\ref{app:low_kicks}). 

\section{Conclusions} Gravitational recoil is a strong-gravity effect of paramount importance in many astrophysical scenarios. Kick-magnitude estimates~\citep{Measured_kick_magnitude_GW190814,Vijay_GWKick} are crucial to understand the retention probability of BHs in their environments and, therefore, their ability to build hierarchical formation channels that can drive the formation of intermediate and supermassive BHs~\citep{GerosaFishbach21}. BHs recoiling through dense environments like Active Galactic Nuclei (AGN) can yield counterpart electromagnetic flares to GW signals~\citep{2019ApJ...884L..50M}, as those recently proposed in~\cite{flare,flares_new}. If real, such multi-messenger observations can enable, {\it e.g.}, independent estimates of the Hubble constant~\cite{Suvodip_H0,H0_HsinYu}. Since the flare properties and observability depend on the direction of the kick w.r.t. both the host AGN and the observer~\cite{ZTF_1,ZTF_2}, obtaining such information from GWs can help to assess the plausibility of the flare as a counterpart and as a probe of the AGN properties. For instance, if the kick is directed away from the observer, the corresponding flare will be obscured by the optically thick AGN disk \footnote{\textcolor{black}{We note that while this only makes use of $\theta_{KN}$, such estimate requires measurements of both the inclination $\iota$ and azimuth $\phi_N$ of the binary.}}. While there is no candidate electromagnetic counterpart to GW190412, we foresee the usage of the recoil direction to assess the plausibility of future candidates.\\ 

Our measurement requires that of the \textit{physically meaningful} azimuthal angle $\phi_N$ of the observer around the source -- misleadingly known as ``coalescence phase'' -- that can be compared to the kick azimuthal angle $\phi_K$. We hope that as gravitational-wave detectors and search techniques improve their respective sensitivities, increasing the number of events with HM content, this will become common practice. \cor{Finally, we note that the restriction to the quasi-circular case imposes constraints on the remnant kick magnitudes. Should GW190412 be consistent with an eccentric binary, larger kicks may be allowed~\cite{Sperhake2020_ecc,Radia2021}.}\\
\appendix
\renewcommand\thesection{\Roman{section}}
\renewcommand\thesubsection{}
\renewcommand\theequation{\Roman{section}.\arabic{equation}}
\section{Bayesian evidences for analyses using restricted kick-magnitude ranges\label{app:BayesZ}}

As discussed in the main text, when considering the full possible range of kick magnitudes, our posterior distribution for the kick magnitude largely follows the prior, leading to a \cor{fairly} unconstrained kick magnitude. Even in this situation, however, we find that the data is informative enough to rule out small kicks below 50\,km/s with high probability. In this Appendix, we describe the procedure followed to obtain this result.\\

\subsection*{Evaluating the probability of a non-zero kick}

Given our priors on mass ratios and spins, the prior probability for a zero kick is exactly zero, which automatically yields a null posterior probability. This is astrophysically sensible, as the conditions required for null kicks are extremely restrictive. For instance, for the case of non-spinning binaries, the kick will only be zero if both black holes have exactly the same mass, which represents a sub-space of zero volume within all the possible mass ratios. In such scenario, however, it may be argued that the inference of a \cor{non-zero} kick may just be driven by prior assumptions and not from information retrieved from the data.\\

In this situation, one can evaluate the need for a non-zero kick to explain the data in a prior-independent way. In particular, the relative probability for a signal model imposing zero and non-zero kicks can be obtained by simply performing two different parameter inference runs -- the one presented in the main text and one imposing $K=0$ -- and computing the ratio of respective Bayesian evidences, ${\cal{Z}}$ and ${\cal{Z}}(K=0)$, known as Bayes Factor ${\cal B}$. The kick magnitude, however, is not an explicit parameter of waveform models but an implicit one determined by the mass ratio and spins of the source, which makes cumbersome to perform the mentioned experiment. In this situation, it is well known that ${\cal{B}}$ can be simply obtained through the \cor{Savage-Dickey density ratio}~\cite{SD-ratio}, equal to the ratio of the prior and posterior probabilities for $K$ evaluated at $K=0$. However, since in our case both prior and posterior equal to zero (see the top panel of Fig.~\ref{fig:ppm}), such calculation can also be problematic.\\

\subsection*{Evaluating the probability for a kick above and below a finite value}

In order to obtain robust results, we choose instead to compute the Bayes Factor ${\cal{B}}^{+}_{-}(K_0)$, \textcolor{black}{defined as the ratio of the evidences ${\cal{Z}}^{+}$ and ${\cal{Z}}^{-}$ respectively corresponding to analyses (or source models) restricted to kick magnitudes $K$ above and below a given threshold $K_0$. This way, both prior and posterior are non-zero in the ranges of interest.} As we show later, this can be simply obtained as 

\ncor{
\begin{equation}
   {\cal{B}}^{+}_{-}(K_0)=\frac{{\cal{Z}}^{+}}{{\cal{Z}}^{-}}=\frac{\int_0^{K_0}\pi(K)\,\dd K}{\int_{K_0}^{\infty}\pi(K)\,\dd K} \times \frac{\int_{K_0}^{\infty}p_{\text{marg}}(K)\,\dd K}{\int_{0}^{K_0}p_{\text{marg}}(K)\,\dd K}.
\label{eq:bayes}
\end{equation}
}

Above, $p_{\text{marg}}$ denotes the posterior probability for $K$ marginalised over all other parameters while $\pi(K)$ denotes the corresponding prior probability. These are respectively the grey and red distributions for $K$ shown in Fig.~\ref{fig:kick_pe}.\\ 

\begin{figure}
\begin{center}
 \includegraphics[width=0.49\textwidth]{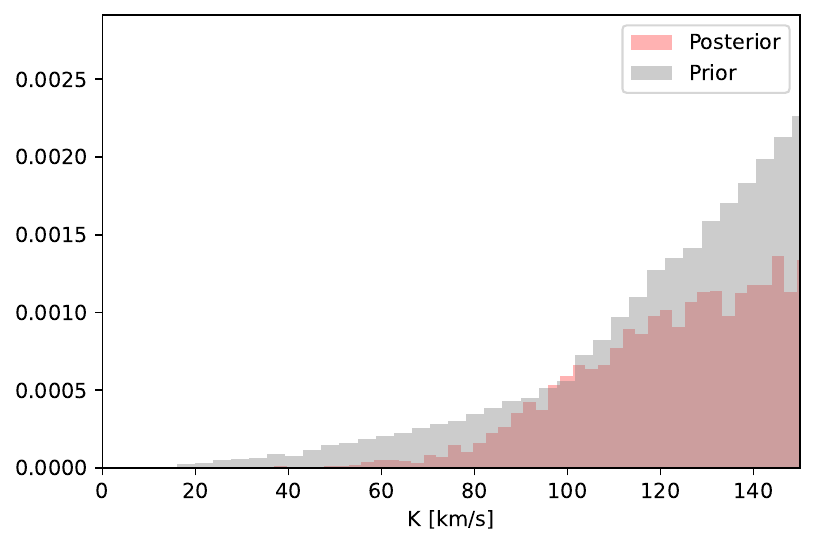}
\includegraphics[width=0.49\textwidth]{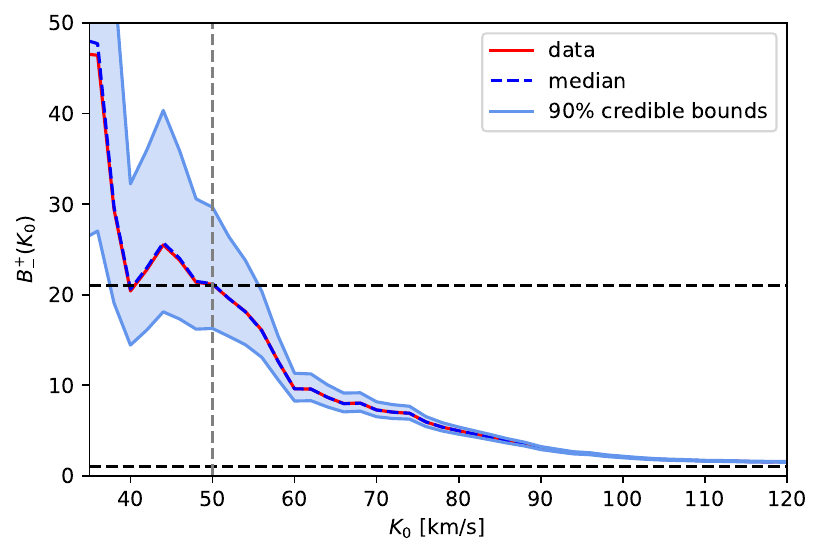}
\caption{\textbf{Discarding small recoil magnitudes}. The top panel shows the prior and posterior distributions for the kick magnitude $K$ of GW190412, zooming in the $K<120$\,km/s region. The bottom panel shows the Bayes Factor ${\cal{B}}^{+}_{-}(K_0)$, between template models restricted to sources above and below a given threshold $K_0$. The red line shows the result obtained from the posterior and prior shown in the top panel. The blue line and contours denote the median and $90\%$ credible bounds obtained through the generation of 500 random re-samplings. Finally, the two horizontal lines denote Bayes factor values of 1 (equal preference) and 21.}
\label{fig:ppm}
\end{center}
\end{figure}

We show the mentioned distributions again in the top panel of Fig.~\ref{fig:ppm}, zooming in the region $K<120$\,km/s, which was imperceptible in Fig.~\ref{fig:kick_pe}. First, it is rather obvious that the posterior deviates from the prior in that region, indicating that the data is indeed informative in the low-K end. Second, while the prior has support all the way to $K=0$, the posterior shows \textcolor{black}{negligible probability for} $K<50$\,km/s, which is the typical escape velocity of globular clusters \textcolor{black}{and young-star clusters}\cite{HolleyBockelmann2008_GlobClusterScape,Merritt:2004xa,Antonini2016,Stoop2023_esc_ygc, Stoop2023_21kms, PortegiesZwart2010,Mapelli2021_escape_vels}. In other words, the data is informative enough to overcome the \textcolor{black}{non-negligible prior probability} for such low kicks. Using Eq. \eqref{eq:bayes}, we obtain \cor{${\cal{B}}^{+}_{-}(50) \simeq 21$. This is, a model restricted to $K>50$\,km/s is $\simeq 21$} times more probable given the data than a model restricted to $K<50$\,km/s. Equivalently, $K>50$\,\textcolor{black}{km/s} with $\sim 95\%$ probability.\\ 

The red line in the bottom panel of Fig.~\ref{fig:ppm} shows ${\cal{B}}^{+}_{-}(K_0)$ as a function of the $K_0$ cutoff. First, we note that, as expected by looking at the top panel, ${\cal{B}}^{+}_{-}$ increases for decreasing $K_0$. The reason is that the ratio between the posterior and the prior \textcolor{black}{probabilities} for $K<K_0$ decreases for decreasing $K_0$. This is, the data increasingly overcomes the (already small) preference of the prior for small kicks below $K_0$ as $K_0$ decreases. On the contrary, for instance, prior and posterior show the same \textcolor{black}{probability} for \cor{$K<450$\,km/s (and, therefore, for $K>450$\,km/s), thus yielding ${{\cal{B}}^{+}_{-}(450)}=1$.}\\

We note that the obtention of ${\cal{B}}^{+}_{-}(K_0)$ is potentially subject to large uncertainties given the small sample count in the low kick region. To estimate such uncertainty, we generate 500 random realisations of the prior and posterior distributions (a process known as bootstrapping) and compute ${\cal{B}}^+_-$ for each of them. The blue region in the bottom panel of Fig.~\ref{fig:ppm} shows the $90\%$ credible intervals of ${\cal{B}}^{+}_{-}(K_0)$ as a function of $K_0$. The dashed blue shows median value. \cor{We obtain ${\cal{B}}^{+}_{-}(50)=21.2^{+8.5}_{-4.9}$ at the $90\%$ level or, equivalently, $p(K>50) = 0.956^{+0.012}_{-0.013}$}. As expected, uncertainties increase \ncor{for decreasing $K_0$, due to the decreasing number of posterior samples, which becomes null for $K_0 < 31.6$\,km/s leading to somewhat artificial values ${\cal{B}}^{+}_{-}(K_0) = \infty$. We may assume, however, that for such $K_0$ values the true lower end of the $90\%$ credible interval should be equal or larger than that for $K_0 < 31.6$\,km/s, which yields ${\cal{B}}^{+}_{-}(31.6) = 70$ }.\\

\cor{Finally, we note that this study ignores the intrinsic error in the estimation of the kick arising in the actual numerical simulations due to, {\it e.g.}, finite grid resolution. We show at the end of Appendix~\ref{app:injections} that these do not seem to induce any significant systematic error in our analysis, or in those of significantly louder signals.}

\subsection*{Derivation of Equation~\eqref{eq:bayes}}

The posterior probability for source parameters $\vec{\theta}$ given  data $d$ is given by

\begin{equation}
    \textcolor{black}{p(\vec{\theta}|d)} = \frac{\pi(\vec{\theta}){\cal{L}}(d|\vec{\theta})}{{\cal{Z}}}
\end{equation}

Here, ${\cal{L}}(d|\vec{\theta})$ denotes the likelihood of the source parameters $\vec{\theta}$ and $\pi(\vec{\theta})$ denotes the prior probability for such parameters. Finally, the normalisation term ${\cal{Z}}$ denotes the Bayesian evidence, given by

\begin{equation}
    {\cal Z} = \int_{\Theta} \pi(\vec{\theta}) {\cal L}(d|\vec{\theta}) \,\dd \vec{\theta},
\end{equation}
where $\Theta$ denotes the parameter space spanned by the parameters $\vec{\theta}$.\\

We can re-express ${\cal Z}$ as a function of a parameter of interest {\it e.g.}, the kick magnitude parameter $K(\vec{\theta})$, as \footnote{\textcolor{black}{This assumes that: a) the function $K(\vec\theta)$ exists for all  $\vec\theta \in  \Theta$. b) $K(\theta)$ is well behaved, {\it i.e.}, not divergent, highly-oscillatory, fractal ... We that note these conditions are also implicit to the obtention of the marginalised posterior distribution as that shown in Fig.~\ref{fig:ppm}, which has also been computed in works like~\cite{Vijay_GWKick,GW190521I}}.}

\ncor{
\begin{equation}
    {\cal{Z}} =  \int_{0}^{\infty} \dd K^{'}  \int_{\Theta_{K^{'}}\subset \Theta} \pi(\vec{\theta}) {\cal{L}}(d|\vec{\theta}) \,\dd \vec{\theta}.
\end{equation}
}

Here, $\Theta_{K^{'}} \subset \Theta$ denotes the sub-parameter space spanned by the parameters $\vec{\theta} \in \Theta$ satisfying  $K(\vec{\theta})=K^{'}$. The second integral above is just the marginalised posterior probability for the kick magnitude $K$ --which we will denote by $p_{\rm marg}(K|d)$-- \textcolor{black}{times the evidence ${\cal{Z}}$}. This is represented in red in Figs. 4 and 5. The corresponding marginal likelihood can be obtained as 
\begin{equation}
   {\cal L}_{\rm marg}(d|K) = {\cal{Z}} p_{\rm marg}(K|d) / \pi(K), 
\end{equation}
where $\pi(K)$ denotes the prior on $K$ \textit{induced by the prior $\pi(\vec\theta)$}. The latter is represented in grey in Figs. 4 and 5 and is given by:

\ncor{
\begin{equation}
    \pi(K') = \int_{\Theta_{K^{'}}\subset \Theta} \pi(\vec{\theta})  \,\dd \vec{\theta}.
\end{equation}
}

With this, we can re-express the evidence ${\cal Z}$ as

\begin{equation}
    {\cal{Z}} =  \int_{0}^{\infty}  \pi(K) {\cal{L}}_{\rm marg}(d|K) \, \dd K.
\end{equation}

\subsubsection*{Evidence for restricted kick ranges}

Let $\pi(K)$ denote the prior for the kick magnitude of our analysis allowing for generic kicks. Next, we need to impose a new prior $\pi^{-}(K)$ that corresponds to a restriction of our generic analysis to $K<K_0$. To do this, we define $\pi^{-}(K)$ as $\pi^{-}(K) = \alpha^{-} \pi(K)$ if $K<K_0$ -- with $\alpha$ a normalisation constant -- and $\pi^{-}(K) = 0$ if $K > K_0$. \\ 

The Bayesian evidence under the new prior $\pi^{-}(K)$ is given by

\begin{equation}
    {\cal{Z}}^{-}  =  \int_{0}^{\infty} \pi^{-}(K) {\cal{L}}_{\rm marg}(d|K) \,\dd K.  
\end{equation}
    Dividing and multiplying by the original prior $\pi(\theta)$, we can re-express the above as
\begin{equation}
   {\cal{Z}}^{-} =  \int_{0}^{\infty}  \pi^{-}(K) \frac{\pi(K)}{\pi(K)}{\cal{L}}_{\text{marg}}(d|K) \,\dd K, \\
\end{equation}
 which we can re-express as 
\begin{equation}
   {\cal{Z}}^{-} = \int_{0}^{\infty}  \frac{\pi^{-}(K)}{\pi(K)} p_{\rm marg}(d|K) \,\dd K\ . 
\label{eq:eq7}
\end{equation}

With this, using our definition for $\pi^{-}(K)$, we can re-express Eq.~\eqref{eq:eq7} as simply

\begin{equation}
\begin{aligned}
    {\cal Z}^{-} = \alpha^{-} \int_{0}^{K_0} p_{\rm marg}(d|K)\,\dd K. 
\end{aligned}
\end{equation}

Finally, since the prior $\pi^{-}(K)$ must satisfy  $\int_{0}^{\infty} \pi^{-}(K)\dd K = \int_{0}^{K_0} \alpha^{-} \pi(K)\dd K = 1$, the normalisation constant $\alpha^{-}$ given by

\begin{equation}
\alpha^{-} =  \frac{\int_0^\infty \pi(K)\,\dd K}{\int_0^{K_0} \pi(K)\,\dd K} .
\end{equation}.

Analogously, it is easy to check that the Bayesian evidence $\cal{Z}^{+}$ for a model constrained to kicks $K > K_0$ is given by

\begin{equation}
\begin{aligned}
    {\cal{Z}}^{+} = \alpha^{+} \int_{K_0}^\infty p_{\rm marg}(d|K) \,\dd K, 
\end{aligned}
\end{equation} 

with 

\begin{equation}
\alpha^{+} = \frac{\int_{0}^\infty \pi(K)\,\dd K}{\int_{K_0}^\infty \pi(K)\,\dd K} .
\end{equation}.

With this, we arrive at the result presented in Eq.~\eqref{eq:bayes}:

\begin{equation}
   {\cal{B}}^{+}_{-}(K_0)=\frac{{\cal Z}^{+}}{{\cal{Z}}^{-}}=\frac{\int_0^{K_0}\pi(\textcolor{black}{K})\,\dd \textcolor{black}{K}}{\int_{K_0}^{\infty}\pi(\textcolor{black}{K})\,\dd \textcolor{black}{K}} \times \frac{\int_{K_0}^{\infty}p_{\rm marg}(K) \,\dd K}{\int_{0}^{K_0}p_{\rm marg}(K) \,\dd K}.
\end{equation}

 \section{Assessment of systematic errors of \texttt{NRSur7dq4Remnant} for low kick magnitudes\label{app:low_kicks}}

The package \texttt{NRSur7dq4Remnant}~\cite{NRSur7dq4} can provide accurate estimates of the magnitude and direction of black-hole recoils. However, the authors noted that, in some cases, direction estimates can carry errors above $10\deg$ with respect to those coming from numerical relativity simulations. In particular, \ncor{Fig.~7 in Ref.~\cite{NRSur7dq4} shows} that such errors are more common when the kick magnitude is below ${\sim} 300$\,km/s.\\

Since our posterior distribution for the kick magnitude shows \textcolor{black}{non-negligible probability} for $K<300$\,km/s, it is reasonable to ask whether the mentioned errors have an impact in on our results, affecting their reliability and robustness. To check that \textit{this is not the case}, we have conducted a series of tests that we describe next.

\subsection*{Systematic errors I: placing restrictions on the kick magnitude}

\begin{figure}
\begin{center}
\includegraphics[width=0.49\textwidth]{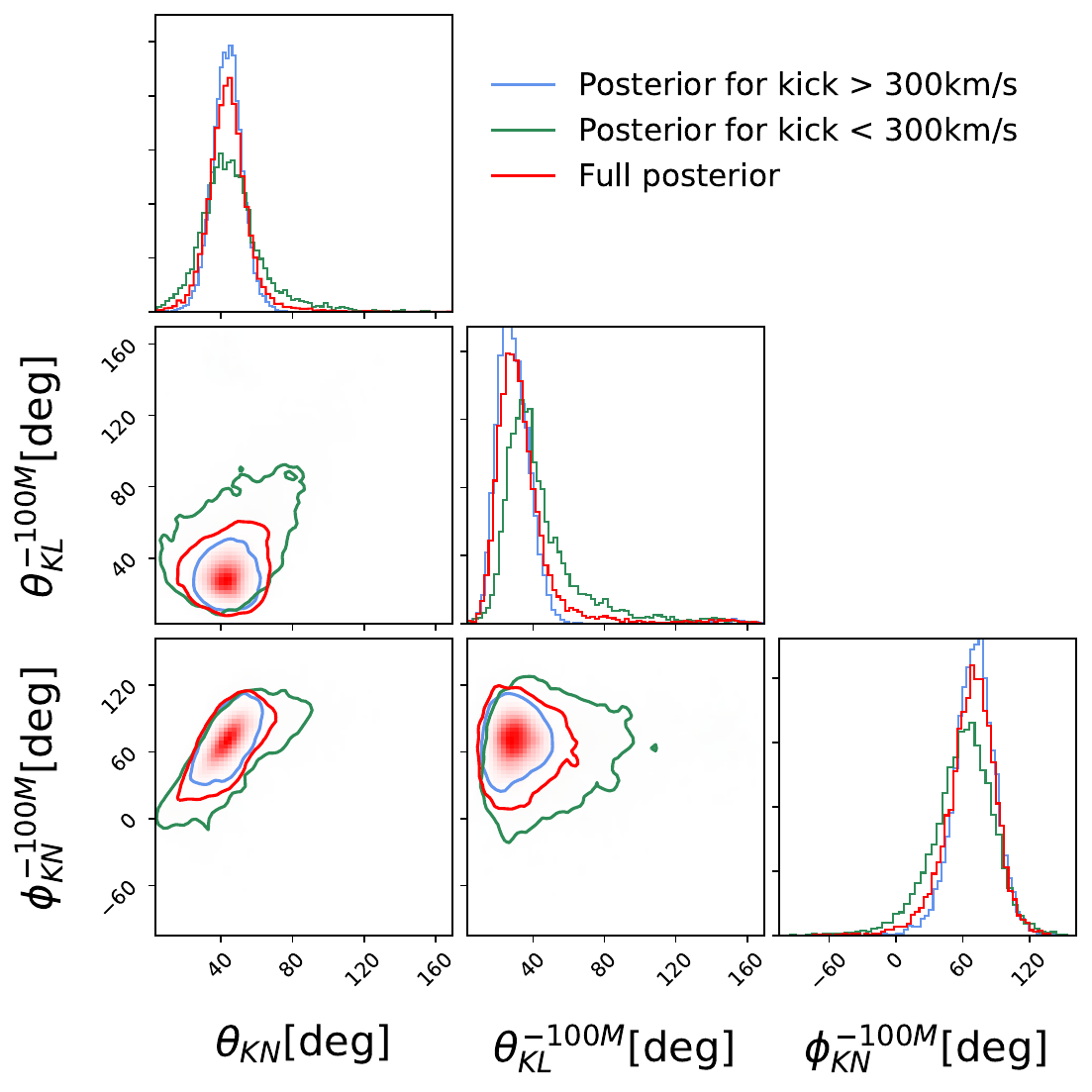}
\caption{\textbf{Comparison between full results and those restricted to kick magnitudes above and below 300\,km/s}. Same as Fig.~\ref{fig:kick_pe} in the main text, but comparing the results from our full inference to those restricting to kick magnitudes above and below 300\,km/s.}
\label{fig:pe_300}
\end{center}
\end{figure}

Figure~\ref{fig:pe_300} shows our one- and two-dimensional posterior distributions for the kick angles \textcolor{black}{of GW190412} (just like Fig.~\ref{fig:kick_pe} in the main text) for the case of our full analysis (red) and for the cases where the kick magnitude is restricted to values above (blue) and below (green) 300\,km/s. The estimate of $\phi_{KN}^{-100M}$ is nearly identical in all cases. The estimate of the angle between the kick and the orbital angular momentum $\theta_{KL}^{-100M}$ is consistent across the different experiments, with those including kicks above 300\,km/s showing a \cor{very slight} stronger preference for low values, {\it i.e.}, for kicks more aligned with the orbital angular momentum. Far from reflecting any systematic error, this is just a consequence of the well-known fact that stronger kicks happen for strongly precessing sources (including superkicks), which tend to kick the final black hole away from the orbital plane. Finally, as a consequence of the previous two results, the estimates for $\theta_{KN}$ are also consistent, with those allowing for larger kick magnitudes being more constrained. 

\subsection*{Systematic errors II: aligned spins vs. random spin direction}

Sources with spins (anti)-aligned with the orbital angular momentum have their kicks constrained below $\sim 500$\,km/s \cite{Healy2014_kick_AS500}. Fig.~\ref{fig:kick_pe} in the main text shows our kick-angle estimates for these sources, as well as those for our generic analysis, allowing for random spin directions. As discussed, the posterior distributions for $\phi_{KN}^{-100M}$, which is the only angle that can be fairly compared, are completely consistent. The distributions for $\theta_{KL}^{-100M}$, however, are necessarily inconsistent because aligned-spin systems have their kick constrained to the orbital plane, yielding a delta distribution at $\theta_{KL}=\pi/2$.

\subsection*{Systematic errors III: back-of-the envelope estimate}

\textcolor{black}{Fig.~3 in Ref.~\cite{SurfinBH} shows the error in the estimation of the angle between the true kick of validation NR simulations and those predicted by the surrogate models \texttt{NRSur3dq8} and \texttt{NRSur7dq2}. These are respectively restricted to aligned-spin binaries up to $q=8$ and precessing binaries up to $q=2$} \footnote{\textcolor{black}{While the same figure is shown in Fig. 7 of \cite{NRSur7dq4} for the case of \texttt{NRSur7dq4}, the much higher density of points therein prevents the ``quantitative'' study in the next paragraph. We therefore use the results in ~\cite{SurfinBH} as a proxy.}}. As one can see therein, in some cases errors are above 10~deg, which is below our typical statistical uncertainties, with the vast majority of them being between 0.02 and 0.1~rad (0.22 and 5~deg). Furthermore, error values clearly accumulate at the low end of the mentioned interval. Importantly, it is also clear that large errors are more common for sources with kicks below 300\,km/s, especially when these have aligned spin (purple dots therein).\\

To make a pessimistic estimate of the potential impact of large errors, we consider a scenario where all our cases below 300\,km/s correspond to spin-aligned sources. With this, we count ${\cal{O}}(20)$ cases above $10 \deg$ and ${\cal{O}}(70)$ cases below such limit, indicating that, in this \textit{unrealistically pessimistic scenario}, only $~30\%$ of such cases would have errors above our typical statistical uncertainty. While we do not know the exact parameters of the validation sources used in the mentioned figure, we can ballpark that since cases with $K < 300$\,km/s represent $~43\%$ of our posterior distribution, then roughly $~13\%$ of our samples would typically suffer from potentially significant systematic errors. In any case, the test above shows that our results are robust against the discussed systematic errors.\\ 

\subsection*{Systematic errors IV: NR resolution}

\begin{figure}
    \centering
    \includegraphics[width=0.45\textwidth]{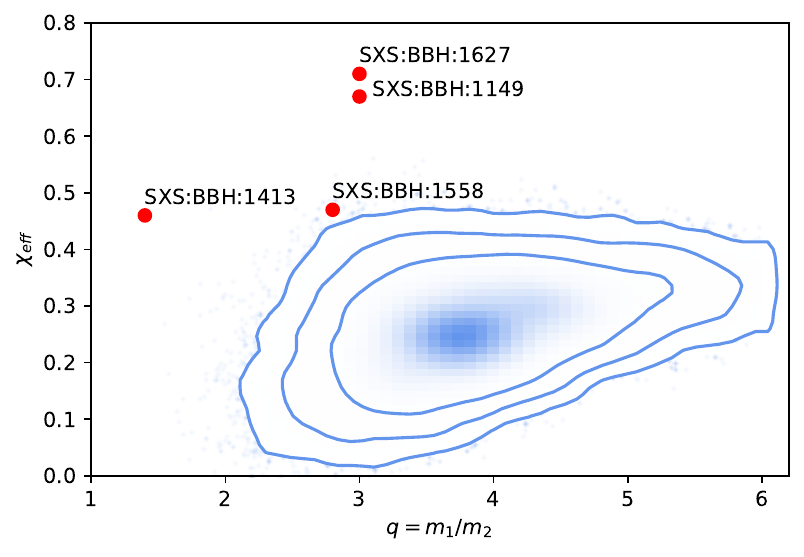}
    \caption{\textbf{Impact of numerical resolution errors in small kick estimates for GW190412.} Two-dimensional $90\%$, $98\%$ and $99.7\%$ posterior credible contours for the mass ratio and effective spin parameters of GW190412. Together, we show the parameters of the 4 publicly available SXS simulations with magnitude recoils smaller than 50\,km/s for which the difference in the recoil magnitude obtained from the two highest available resolutions exceeds 5\,km/s.}
    \label{fig:nr_syst}
\end{figure}

\textcolor{black}{We study in detail the potential impact that errors in the estimations of the kick magnitudes arising from the finite resolution of NR simulations can have on our estimate of the probability that the GW190412 kick is below 50\,km/s. We note that, unfortunately, we could not estimate this from the Fig.~3 in Ref.~\cite{SurfinBH} mentioned above or Fig.~7 in ~\cite{NRSur7dq4}, as those do not report numerical errors for the kick magnitude a function of the kick magnitude itself. To tackle this, we compute the kick magnitude for all of the simulations available in the SXS catalogue for which the kick is below $50$\,km/s, for the two highest levels of accuracy that are publicly available. Out of the 47 simulations we have studied, only 4 show differences larger than $5$\,km/s. These cases show errors of $30$, $15$, $10$ and $6$ km/s, with the first two being clear outliers of a distribution where $90\%$ and $95\%$ of the cases are respectively below 3\,km/s and 5\,km/s. Moreover, Fig.~\ref{fig:nr_syst} shows that the 4 mentioned simulations are outside the parameter space of GW190412, as these have either a mass ratio $q < 2$ and/or an effective-spin parameter $\chi_{\rm eff} > 0.46$. We note that since not all of the waveforms used to train \texttt{NRSur7dq4} are publicly available. We believe, however, such waveforms would be of at least equal quality (if not higher) than the ones we have analysed. For instance, we note that in all the mentioned outliers the maximum publicly available resolution level is \texttt{Lev3}, as compared to the maximum resolution of \texttt{Lev6} available for many waveforms in the catalogue. This makes us confident that our qualitative conclusions about the kick magnitude of GW190412 should not be affected by the resolution-error budget of the NR waveforms used to train \texttt{NRSur7dq4}.}\\

\textcolor{black}{Finally, we note that Fig.~7 in~\cite{NRSur7dq4} shows that uncertainties in the kick direction due to NR resolution are comparable to those of those of the \texttt{NRSur7dq4Remnant} model itself, which we have already discussed.}

\subsection*{Full Bayesian Analysis and Bayesian Priors}
\label{app:injections}
In addition to the purely systematic errors discussed above, in Ref.~\cite{Vijay_kick}, the authors tried to estimate the kick of selected BBH mergers \textcolor{black}{simulated with \texttt{NRSur7dq4}} through a full Bayesian analysis. To this, the authors injected numerically simulated signals in \textcolor{black}{simulated Gaussian detector noise} and recovered them with the strategy followed here. In Fig.~4 therein, however, they observe that, for some cases where the kick is below 300\,km/s, the estimate of the direction is significantly biased with respect to the true value. This happens even when the injected signal is extremely loud.\\

First, we note that such analysis is subject to \cor{several} types of ``errors'' or  ``uncertainties'': first, the potential systematic errors discussed above and, second, that of Bayesian priors. In particular, we note that \cor{the} impact of priors is inherent to any analysis and would just indicate that the effect produced by the kicks (or rather, by the source properties leading to such kicks) in the waveforms is not strong enough for to overcome prior assumptions. Therefore, the fact that parameter inference analyses fail to recover true values is not necessarily indicative of systematic errors.

Finally, we also note that there is a potentially worrying case where a strong bias is found in the kick direction, where the injected signal has a signal-to-noise ratio (SNR) of 66. In principle, for such high SNR, one may expect the data to be informative enough to overcome any prior assumptions. \textcolor{black}{However}, there is a chance that such bias may be due to an extreme Gaussian noise realisation. Nevertheless, we note that this corresponds to 1 out of 6 cases where the authors performed injections with kicks around or below 300\,km/s. \textcolor{black}{Moreover, if this is not sourced by an extreme noise realisation, this seems consistent with the fact that, as discussed in Test III above, large systematic errors above $10\deg$ in the \texttt{NRSur7dq4Remnant} estimates are only found in a small fraction of the cases with kicks below 300\,km/s (if one looks at the generic brown cases shown in Fig.~3 of \cite{SurfinBH}, which seems roughly consistent with that for \texttt{NRSur7dq4Remnant} in Fig.~7 of \cite{NRSur7dq4}).}

\section{Bayesian parameter recovery of numerically simulated signals}
\setlength{\tabcolsep}{10pt} 
\renewcommand{\arraystretch}{1.5} 
\begin{table*}[t]
    \sisetup{
    table-number-alignment = center, 
   }
    \centering
    \begin{tabular}{c|c c c c c S[table-format=4.2]}
        SXS code & $Q = m_1 / m_2$  & $a_{1}$ & $a_{2}$ & $\chi_{\rm eff}$ & $\chi_{\rm p}$ & {$v_{\rm k}\,/\,{\rm km\,s}^{-1}$} \\
        \hline
        SXS:BBH:1443 & 5.681 & 0.4079 & 0.7372 & 0.2365 & 0.0000 &    73.95 \\
        SXS:BBH:1156 & 4.387 & 0.4663 & 0.7677 & 0.3299 & 0.2719 &   131.90 \\
        SXS:BBH:0283 & 3.000 & 0.3000 & 0.2999 & 0.3000 & 0.0000 &   113.27 \\
        SXS:BBH:1937 & 4.000 & 0.4001 & 0.0001 & 0.3200 & 0.0000 &    86.16 \\
         \textcolor{black}{SXS:BBH:1593} & 3.500 & 0.7213 & 0.7588 & 0.2531 & 0.6866 &  1040.86 \\
        SXS:BBH:1805 & 3.415 & 0.4849 & 0.7130 & 0.3743 & 0.2215 &   509.16 \\
        SXS:BBH:1676 & 3.253 & 0.4856 & 0.4018 & 0.3841 & 0.2245 &   179.47 \\
        SXS:BBH:1410 & 4.000 & 0.4680 & 0.4647 & 0.2525 & 0.4000 &   301.85 \\
    \end{tabular}
    \caption{Summary of the SXS simulations chosen for our injection study. These have mass ratio $Q$ and effective inspiral-spin parameter $\chi_{\rm eff}$~\cite{Ajith:2009bn,Santamaria:2010yb} roughly consistent with those of event GW190412, together with varying values for the effective-spin precession parameter $\chi_{\rm p}$~\cite{Hannam:2013oca}, with zero characterising non-precessing systems. The last column shows the magnitude of the recoil velocity for each simulation, computed by integrating the momentum flux \cite{Gonzalez:2007hi,Ruiz2008:RecoilFluxes}.}
    \label{tab:SXS_params}
\end{table*}

\begin{table}[h]
    \sisetup{
    table-number-alignment = center, 
   }
    \centering
    \begin{tabular}{rl|c }
        \multicolumn{2}{c|}{Parameter} & True value \\
        \hline
        RA & $\alpha$ & \SI{3.818}{\radian} \\
        DEC & $\delta$ & \SI{0.640}{\radian} \\
        Polarisation angle & $\psi$ & \SI{2.478}{\radian} \\
        Geocentric time & $t_c$ & \SI{1239082262.18}{\s}
    \end{tabular}
    \caption{True values of the sky-localization and polarisation parameters used for the SXS injection study. These correspond to the maximum likelihood values obtained for GW190412. }
    \label{tab:SXS_extrin_params}
\end{table}

\begin{figure*}
    \centering
    \includegraphics[width=0.77\textwidth]{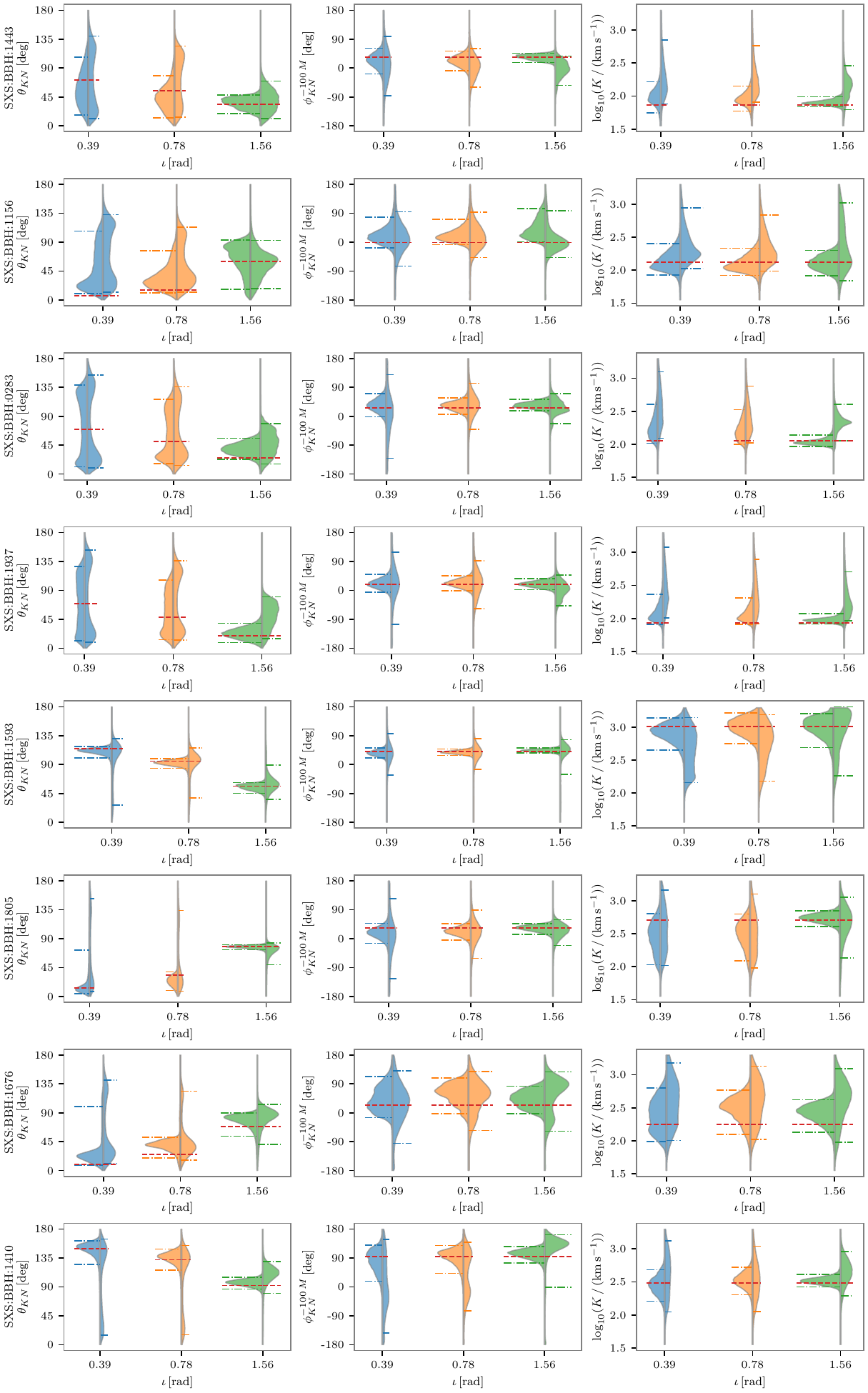}
    \caption{Posterior distributions for the kick parameters, $\theta_{KN}$, $\phi_{KN}^{-100M}$, and magnitude $K$ resulting from the analysis of numerical simulations computed by the SXS code, injected in zero noise. The left/right sides of each violin respectively correspond to results with different SNRs of 50 and 18. Each row of panels corresponds to a different source, roughly consistent with GW190412. We summarise the intrinsic parameters of our injections in Table~\ref{tab:SXS_params}. Sky-location and polarisation angle are described in Table~\ref{tab:SXS_extrin_params}. Blue, orange and green violins correspond to true source inclinations of 0.3, 0.78 and 1.56 radians. \ncor{The red dashed lines denote the true parameter values, which we obtain by  integrating momentum fluxes of the SXS simulation~\cite{Gonzalez:2007hi,Ruiz2008:RecoilFluxes}.} The other coloured dashed lines enclose the symmetric 90\% credible intervals around the median.}
    \label{fig:inj_results}
\end{figure*}

\textcolor{black}{Finally, we have performed a full parameter inference study on numerically simulated signals to further confirm the robustness of our results. We have injected in zero-noise simulated numerical-relativity signals for sources with parameters roughly consistent with those of GW190412}, \textcolor{black}{all obtained from the SXS waveform catalog~\cite{SXS,Boyle2019_SXS,Mrou2013_SXS,Chu2016_SXS0283,Varma2019_SXS1443,NRSur7dq4}. \textcolor{black}{These include all modes up to $|\ell| \leq 4$ and are extrapolated to null infinity using an $\rm N =2$ polynomial~\cite{Boyle2019_SXS}}. We recover these signals using the \texttt{NRSur7dq4} waveform model, estimating the corresponding final parameters using the \texttt{NRSur7dq4Remnant} model. We use the same exact detector configuration, power-spectral densities and Bayesian priors as in the main text. Table~\ref{tab:SXS_params} describes the intrinsic parameters of the numerical-relativity simulations we selected \textcolor{black}{while Table~\ref{tab:SXS_extrin_params} describes the chosen sky-location and polarisation}. \textcolor{black}{Three} of the systems have aligned spins, while the other \ncor{five} are precessing. \textcolor{black}{Our injections span a wide range of kick magnitudes and expand outside the mass-ratio training range of \texttt{NRSur7dq4}. The latter is motivated by the fact that the posterior distribution for this parameter for GW190412 peaks around $q\sim 4$, where the end of the training range sits. Also, we note that while the posterior for the secondary spin shows \textcolor{black}{non-zero probability} for $a_2 > 0.8$ -- also surpassing the limits of the surrogate training region -- such posterior is rather uninformative, and any systematics should not impact our results. In fact, we have checked that visually identical posterior distributions are obtained for all kick parameters \textcolor{black}{of GW190412} if we restrict to $a_2 < 0.8$.} GW190412 has an intermediate source inclination of $\theta_{JN} = 0.7 \pm 0.2$ measured at a reference frequency of 20\,Hz~\cite{GW190412}. For the sake of completeness we have selected three different source inclinations $\iota \in \{0.39, 0.78, 1.56\}$\,rad specified at $t=-100\,M$. We note that while \textcolor{black}{$\theta_{JN}$ and $\iota$} define different notions of inclination for precessing sources, the broad inclination range that our injections cover contains well those consistent with GW190412. Combining this choice of $\iota$ with a rather \textcolor{black}{arbitrary choice of $\phi_N^{-100M}=6.1$}, our injections cover a reasonably wide range for the kick angles $\theta_{KN}$ and $\phi_{KN}^{-100M}$. In addition, for each injection we have chosen two distances such that the SNR of the signals is 18 and 50, respectively, around and significantly above that of GW190412. The total \textcolor{black}{detector-frame} mass is set to be $M = \num{45.76}\,M_\odot$ consistent with that of GW190412. We used the same sampling setting as in the main text, with 4096 live points.}\\

Figure~\ref{fig:inj_results} shows the posterior distributions, together with the $90\%$ credible intervals and the corresponding true values for the kick parameters\footnote{\textcolor{black}{We compute the true values by explicitly performing the corresponding momentum integral (see {\it e.g.}~\cite{Gonzalez:2006md}) instead of reading them from the metadata of the simulations, as the latter corresponds to a coordinate velocity~\cite{Boyle2019_SXS}}}. As a general rule, the true values are always within the $90\%$ credible intervals even for SNRs of 50, way beyond that of GW190412. \textcolor{black}{There are only two exceptions to this, out of the 144 posteriors we show. The first one corresponds to the posterior} for $\theta_{KN}$ obtained for the \texttt{SXS:BBH:1156} case for the lowest inclination we consider, for both SNRs. This is, however, due to the extremely small prior \textcolor{black}{probability} for the true value. This is consistent with the fact that the posterior, while biased, moves towards the true value when we raise the SNR, following the peak of the likelihood. \textcolor{black}{The same behaviour can be observed for the \texttt{SXS:BBH:1937} cases with $\iota=0.29$ and $\iota = 1.56$}. We note that this is something that all GW analysis are subject to independently of the accuracy of the waveform models and would therefore happen even if the \texttt{NRSur7dq4} was infinitely accurate. \textcolor{black}{The second case corresponds to the $\phi_{KN}^{-100M}$ posterior for the \texttt{SXS:BBH:1156} case for ${\rm SNR} = 50$ and $\iota=1.56$, for which the true value is slightly away of the $90\%$ credible interval while being well inside it for the ${\rm SNR} = 18$ case. In this case, we have evidence this is caused by actual waveform systematics that make \texttt{NRSur7dq4} waveforms with biased values of $\phi_{KN}^{-100M}$ fit the SXS waveform better than predicted for the true injected parameters. In particular, we have checked that the posteriors for ${\rm SNR} = 50$ match well the restriction of those for ${\rm SNR} = 18$ to only high likelihood parameters. Nevertheless, we note that consistently with \cite{Islam2021_GW190412}  our results rule out an edge-on inclination for GW190412 and stress that its SNR is rather around 18, for which we obtained unbiased estimates.}\\

\textcolor{black}{We note that the analysis of aligned-spin cases, while unbiased, proved to be particularly challenging. The reason is that the null prior \textcolor{black}{probability} for aligned-spins made sampler convergence particularly cumbersome, requiring us to use significantly more aggressive sampler configurations with a larger number of parallel chains than the default ones we used for generic-spin cases \footnote{\textcolor{black}{In particular, for precessing cases we used 4096 live points, 2 parallel chains, $\rm{nact} = 35$ (with $\texttt{rwalk}$) and $\rm{maxmcmc} = 15000$, while for aligned-spin injections we performed analyses using both 8 and 10 parallel chains. The latter two analyses show consistent results, which makes us confident that the runs are well converged.}}. \textcolor{black}{In fact, the analysis of these injections returned Bayes Factors of order $10^4$ favouring aligned spins, indicating indeed that the sampler would need to explore in detail the narrow aligned-spin region of the parameter space. In contrast, GW190412 returns a Bayes Factors of only 2.1 in favour of the aligned-spins, making us confident that such aggressive sampler settings are not needed to analyze it.}}\\

\cor{All in all, our results indicate that our analysis is not impacted by systematic errors coming from waveform modelling. Moreover, this would still be the case even if GW190412 had an SNR of 50, way beyond that of any GW observation to date.}

\section*{Acknowledgements} We thank Nicolas Sanchis-Gual, Barry McKernan and Saavik Ford for their comments on the manuscript. We thank Sergei Ossokine and Nathan Johnson-McDaniel for their advice to perform parameter estimation at a given reference time $t_{\rm ref}$ and Vijay Varma for discussions on the importance of this choice. We also thank Angela Borchers for useful discussions regarding the accuracy of kick estimations of various waveform models. Finally, we thank Thomas Dent, Tjonnie Li and Titus M\"{o}mbacher for useful discussions. The analysed data and the corresponding power spectral densities are publicly available at the online \cor{Gravitational Wave Open Science Center \citep{Abbott2023_Data_GWTC3}}. JCB is supported by a fellowship from ``la Caixa'' Foundation (ID100010434) and from the European Union’s Horizon2020 research and innovation programme under the Marie Skłodowska-Curie grant agreement No 847648. The fellowship code is LCF/BQ/PI20/11760016. JCB is also supported by the research grant PID2020-118635GB-I00 and by a Ram\'{o}n y Cajal Fellowship RYC2022-036203-I from the Spain-Ministerio de Ciencia e Innovaci\'{o}n. KC acknowledges the generous support provided through NSF grant numbers PHY-2207638, AST-2307147, PHY-2308886, and PHY-2309064. We acknowledge using the IUCAA LDG cluster Sarathi for the computational/numerical work. The authors acknowledge computational resources provided by the CIT cluster of the LIGO Laboratory and supported by National Science Foundation Grants PHY-0757058 and PHY0823459, and the support of the NSF CIT cluster for the provision of computational resources for our parameter inference runs. This material is based upon work supported by NSF's LIGO Laboratory which is a major facility fully funded by the National Science Foundation. The authors acknowledge the use of computing facilities supported by grants from the Croucher Innovation Award from the Croucher Foundation Hong Kong. This research has made use of data or software obtained from the Gravitational Wave Open Science Center (gwosc.org), a service of the LIGO Scientific Collaboration, the Virgo Collaboration, and KAGRA. This material is based upon work supported by NSF's LIGO Laboratory which is a major facility fully funded by the National Science Foundation, as well as the Science and Technology Facilities Council (STFC) of the United Kingdom, the Max-Planck-Society (MPS), and the State of Niedersachsen/Germany for support of the construction of Advanced LIGO and construction and operation of the GEO600 detector. Additional support for Advanced LIGO was provided by the Australian Research Council. Virgo is funded, through the European Gravitational Observatory (EGO), by the French Centre National de Recherche Scientifique (CNRS), the Italian Istituto Nazionale di Fisica Nucleare (INFN) and the Dutch Nikhef, with contributions by institutions from Belgium, Germany, Greece, Hungary, Ireland, Japan, Monaco, Poland, Portugal, Spain. KAGRA is supported by Ministry of Education, Culture, Sports, Science and Technology (MEXT), Japan Society for the Promotion of Science (JSPS) in Japan; National Research Foundation (NRF) and Ministry of Science and ICT (MSIT) in Korea; Academia Sinica (AS) and National Science and Technology Council (NSTC) in Taiwan. This manuscript has LIGO DCC number P2200332.

\bibliography{ZTF_Flare.bib}
\bibliographystyle{apsrev4-2}

\end{document}